\documentstyle[11pt,newpasp,twoside,epsf,psfig]{article}
\def\edcomment#1{\iffalse\marginpar{\raggedright\sl#1\/}\else\relax\fi}
\reversemarginpar

\begin{document}
\title{The Peculiar Galaxy NGC~7332}
\author{Jes\'us Falc\'on-Barroso}
\affil{School of Physics \& Astronomy. University of Nottingham,
 University Park, Nottingham, NG7 2RD, United Kingdom.}
\author{Reynier F. Peletier}
\affil{CRAL, Observatoire de Lyon, 9, Av. Charles Andr\'e, 69230 Saint-Genis Laval,
 France \& 
 \\
 School of Physics \& Astronomy. University of Nottingham,
 University Park, Nottingham, NG7 2RD, United Kingdom}

\begin{abstract}
We present a detailed study of the edge-on S0 galaxy NGC~7332. We show that this  object
lies significantly below the fundamental plane of early-type galaxies. To understand this
behaviour, we have  carried out observations with the Integral Field Spectrograph SAURON, which give us
a detailed view of the internal kinematics and gas and stars in this galaxy.
\end{abstract}

\section{Introduction}
NGC~7332 is an ordinary-looking galaxy that is peculiar in many aspects. A
detailed  study of its kinematics of gas, stars and its surface photometry
has shown that it is a Rosetta stone that can teach us what the components are
that galaxies  consist of, and how they form.

The galaxy is a highly inclinated S0 located close to the spiral galaxy NGC~7339
at 5.2$'$ with a similar systemic velocity. Previous papers on this galaxy 
include those by Fisher, Illingworth, \& Franx (1994) and Plana \& Boulesteix
1996). While the main results of Fisher et al. include the B/D decomposition of 
the galaxy, as well as an analysis of its stellar and gaseous kinematics, Plana 
\& Boulesteix focus exclusively on determining the gas morphology of the galaxy 
from H$\alpha$ emission maps. Fisher et al. show that the galaxy has kinematics 
fairly typical of a stellar disk, with a counter-rotating and a fainter co-rotating 
gas disk. An extension of this result was found by Plana \& Boulesteix who not 
only mapped the distribution of the 2 gaseous components, but also modelled them, 
finding different inclination angles for each component. Broad band colours (Balcells 
\& Peletier 1994) show that the colours of NGC~7332 ($B-R=1.4$) are somewhat bluer 
than colours of elliptical galaxies of the same luminosity. This is confirmed by 
a spectral analysis showing that the galaxy has a luminosity weighted age of about 
6 Gyr (Vazdekis \& Arimoto 1999). 

Here we present 2 studies aimed at understanding the difference between NGC~7332 
and ordinary red S0 galaxies. First we present a new study of the position of this
galaxy on the fundamental plane (FP) of early-type galaxies (for details see 
Falc\'on-Barroso et al. 2002). We show that NGC 7332 lies below the relation, 
and that the displacement can not explained by uncertainties in our measurements. 
Second, we show new 2D maps of the stellar and the gaseous kinematics in the [OIII] 
5007 \AA\ line obtained with SAURON.

\section{NGC~7332 on the Fundamental Plane of Early-type galaxies}
We have studied the position of 19 galactic bulges (the sample of Peletier et al. 1999) 
on the FP in both B and K, with the aim of finding out whether bulges are formed in the same way as
ellipticals. We find that bulges are slightly shifted with respect to 
the FP defined by ellipticals. This is explained easily by the fact that the sample
consists of galaxies with inclination $>$ 50$^{\rm o}$, together with the contribution 
of the rotating disk, which causes the velocity dispersion of a bulge to be slightly lower
than an elliptical of the same mass. NGC~7332, however, lies significantly further below 
the FP. Not only is this the case in the B-band, where such a deviation can be 
explained if the galaxy is much younger than other bulges, but also in K,
where young stellar populations cannot cause such a deviation. At this stage of 
our analysis we do not have a satisfactory reason why this galaxy is departing so strongly 
from the FP relation. NGC~7332 is the only S0 galaxy that we know which deviates so 
strongly, without being obviously interacting (see e.g. Schweizer \& Seitzer 1992). 
Another peculiarity of this galaxy is that the stellar colours in this galaxy are so
homogeneous, with only a very small blueing towards the outer parts (Balcells \& Peletier
1994; Fisher et al. 1994). 

In Fig.~1 we show the location of our sample of bulges on the FP of early-type galaxies as
defined by J\o rgensen, Franx \& Kj\ae rgaard 1996 (in the B-band, top panel) and Pahre et al
1998 (in K-band, bottom panel). We have marked NGC~7332 with an open
circle.

\section{Stellar Kinematics}
NGC~7332's kinematics reveals a smooth uniform stellar velocity field rotating along the 
minor axis of the galaxy (Fig.~2). The velocity field appears consistent with cylindrical
rotation, although this result will have to be confirmed with some detailed modeling.
The velocity dispersion map shows  a dip along the major axis of the galaxy (Fig.~3).
This is usually the case when a galaxy has an inner disk. Indeed, for NGC~7332 
such an inner disk in the first $\pm$ 5 arcsec has been discovered by Seifert \& Scorza 
(1996). The  higher order moments Gauss-Hermite moments(h$_3$, h$_4$, van der Marel \& Franx
1993) seem to be regular, which means that it is unlikely that the galaxy harbours multiple
stellar disks, like e.g. in NGC~4550 (Rubin et al. 1992).

\section{Measurement of the emission lines}

In order to obtain the kinematics of the gas, we need to separate emission from absorption
lines. To do this we used a small variation to a more sophisticated technique 
(see de Zeeuw et al. 2002) and subtracted the stellar spectrum
by fitting the observed stellar energy distribution outside the region of the 
emission lines with synthetic templates from the models of Vazdekis (1999)
for composite single age/metallicity models of various ages and metallicities. 

This procedure can be summarize in a few steps:
\begin{itemize}
\item{We first derive the stellar kinematics ($\gamma,V,\sigma$) using the method
of van der Marel \& Franx (1993) with a single stellar population (SSP) model as template.}
\item{Second, we build a library of models with different ages and metallicities (SSP
models from Vazdekis 1999).}
\item{We convolve the spectra of the library with the LOSVDs obtained in step 1.}
\item{We then fit a linear combination of these SSP models to the galaxy spectrum.}
\item{Finally we substract the fitted spectrum from the observed SED to obtain 
a pure emission-line spectrum.}
\end{itemize}

Using this method it is possible to recover faint emission line profiles, as seen 
at the  NI doublet at 5199 \AA (log $\lambda \approx$3.718) in Figures (8 \& 9).

\section{Counter-rotating gas in NGC~7332}
This is without doubt the most unusual feature in NGC~7332. A close inspection
of the [OIII] (5007 \AA) gas emission lines shows a primary, counter-rotating
gas disk  (Figs.~4 \& 5) and a much fainter secondary component (Figs.~6 \& 7) 
corotating with respect to the stellar component. This double component has
been detected also by Plana \& Boulesteix (1996) from H$\alpha$ observations. In
general, both  Plana and we find the same spatial distribution of the second
component. The spatial distribution of the emission in [OIII], however, is not
the same as  that of H$\alpha$.  While we find very little OIII emission in the
center of the galaxy,  Plana \& Boulesteix find a considerable amount of
H$\alpha$.  Trying to understand why this is the case,  we have analysed
long-slit data, in the H$\alpha$ region, from the ING (Isaac Newton Group) 
archive. These data, obtained in June 1994 by Merrifield \& Kuijken, reveal 
that H$\alpha$ in the center is weak contrary to the results of Plana \&
Boulesteix.  This is confirmed by the fact that H$\beta$ in our SAURON
data is weak as well, casting doubt upon the results of Plana \& Boulesteix.
Further analysis about this subject is due to appear in a forthcoming paper.

\section{Conclusions}
\begin{itemize}
\item{NGC~7332 is an unusual object.}
\item{As yet it is a miracle why this galaxy lies so far
away from the fundamental plane of early-type galaxies}
\item{Its stellar kinematics is not peculiar at all.}
\item{The existence of two gaseous components coexisting with opposite angular 
momentum leaves the hypothesis open of a recent merger scenario in
NGC~7332.}
\end{itemize}

\section*{Acknowledgements}

We thank the SAURON team for obtaining and reducing the data of NGC~7332. 
The William Herschel Telescope is operated on the island of La Palma by the
Isaac Newton Group in the Spanish Observatorio del Roque de los Muchachos of
the Instituto de Astrof\'\i sica de Canarias.

\begin{figure}
\begin{center}
\psfig{figure=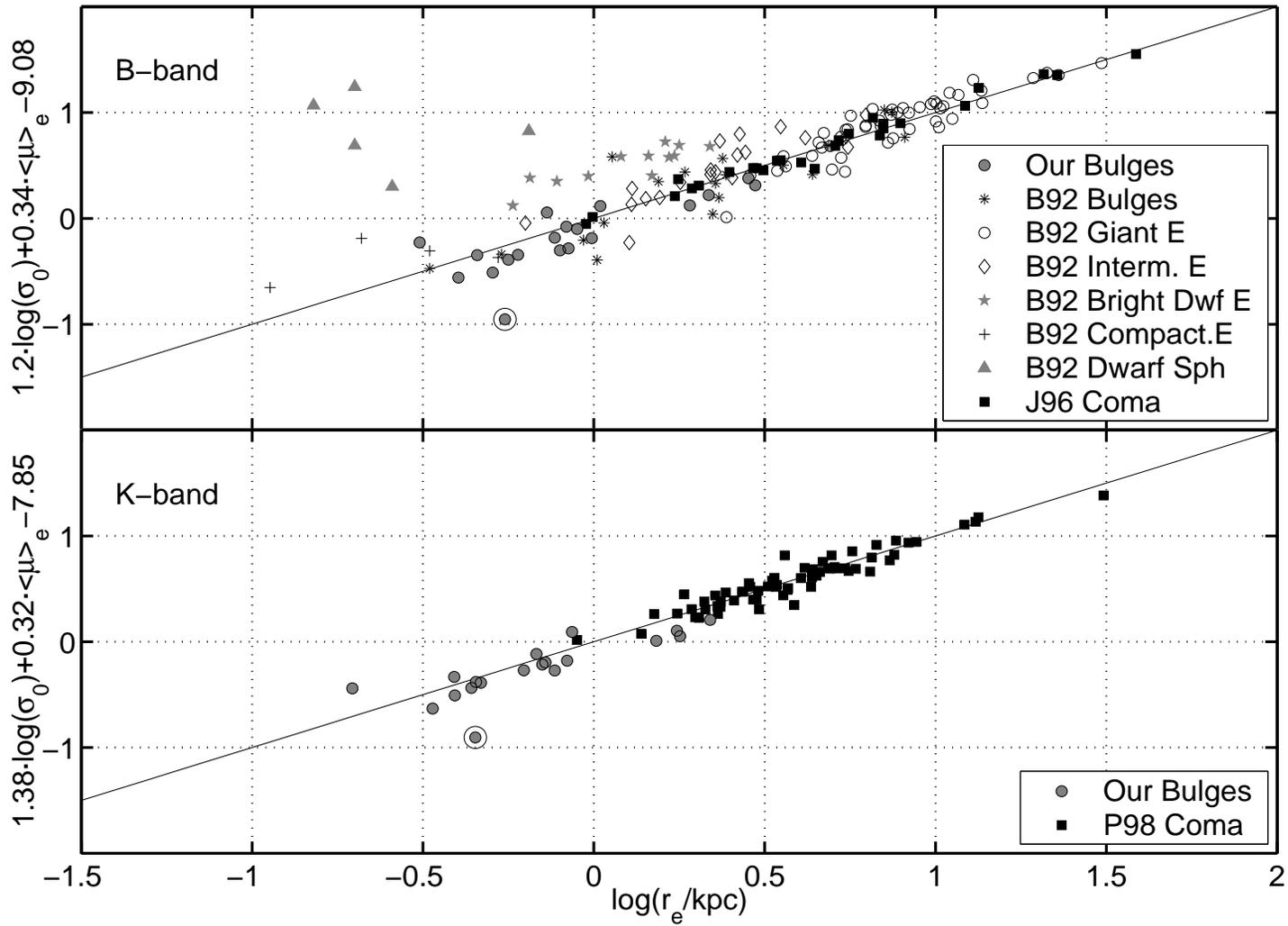,angle=180,height=7.5in,width=5.5in}
\caption{NGC~7332 on the fundamental plane of early-type galaxies.}
\end{center}
\end{figure}

\begin{figure}
\begin{center}
\hspace{5cm}
\psfig{figure=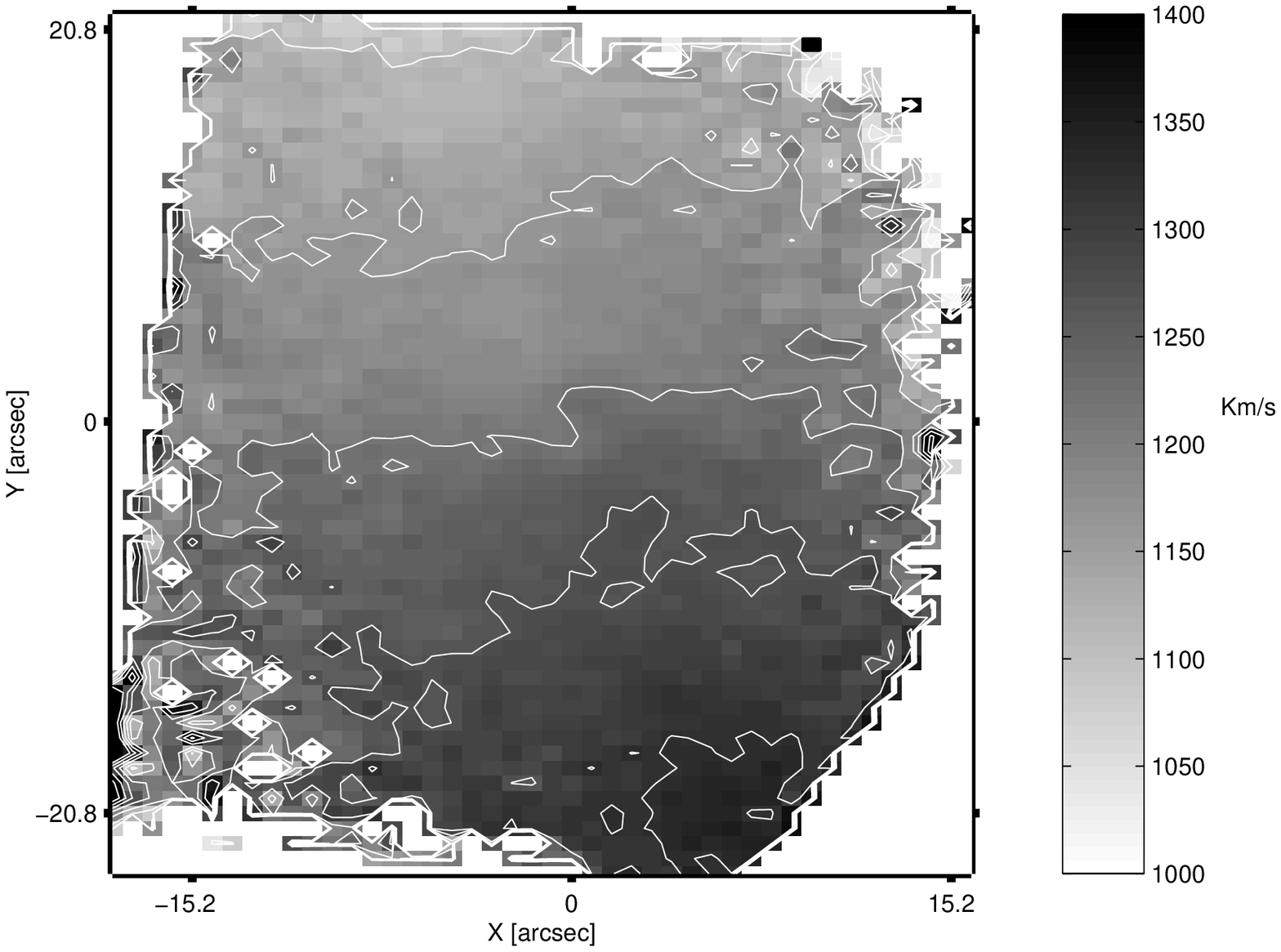,angle=0,height=3.2in,width=4in}
\caption{Stellar velocity map of NGC~7332.}
\end{center}
\end{figure}

\begin{figure}
\begin{center}
\hspace{5cm}
\psfig{figure=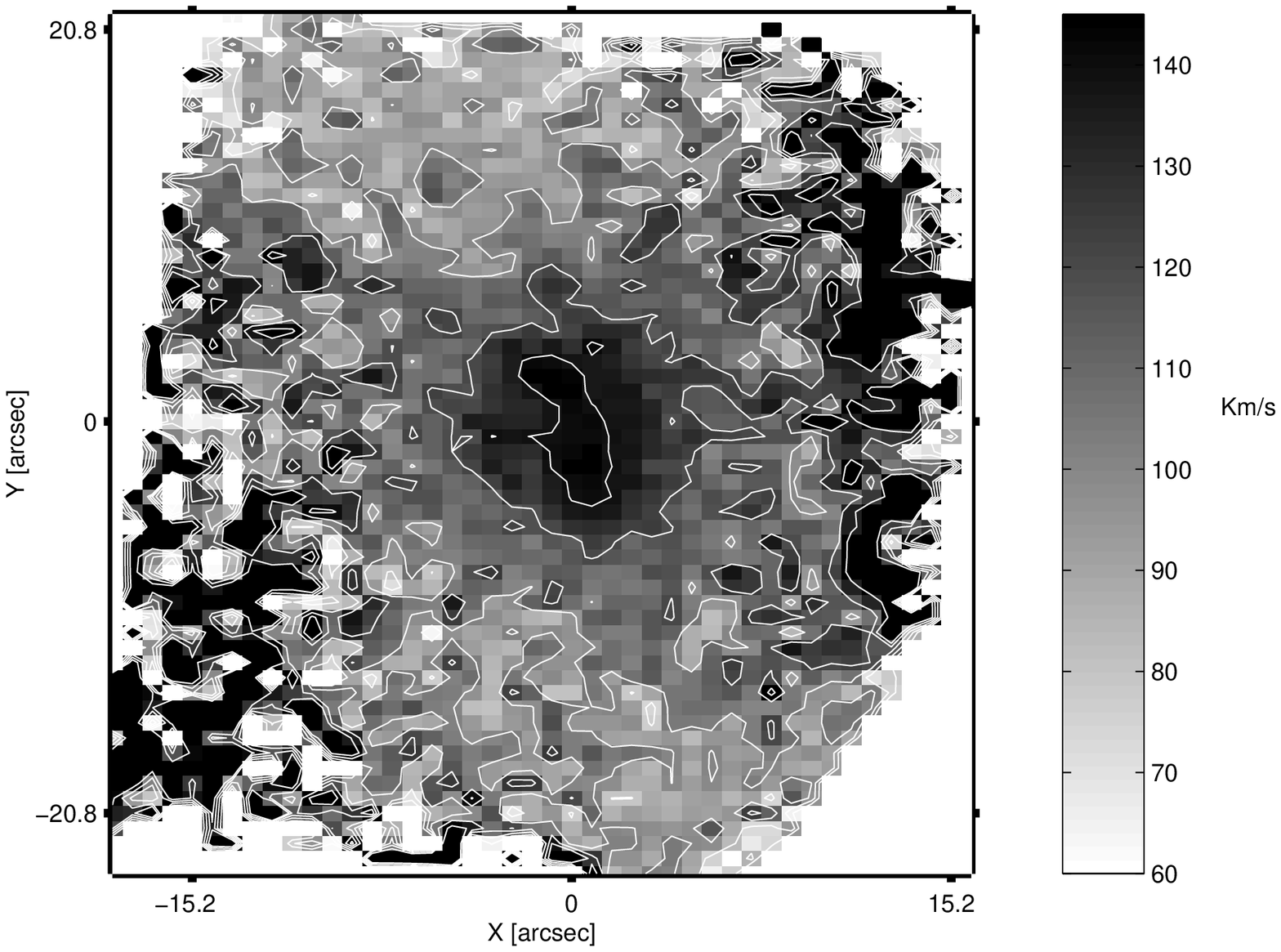,angle=0,height=3.2in,width=4in}
\caption{Stellar velocity dispersion map of NGC~7332.}
\end{center}
\end{figure}

\begin{figure}
\begin{center}
\hspace{-1cm}
\psfig{figure=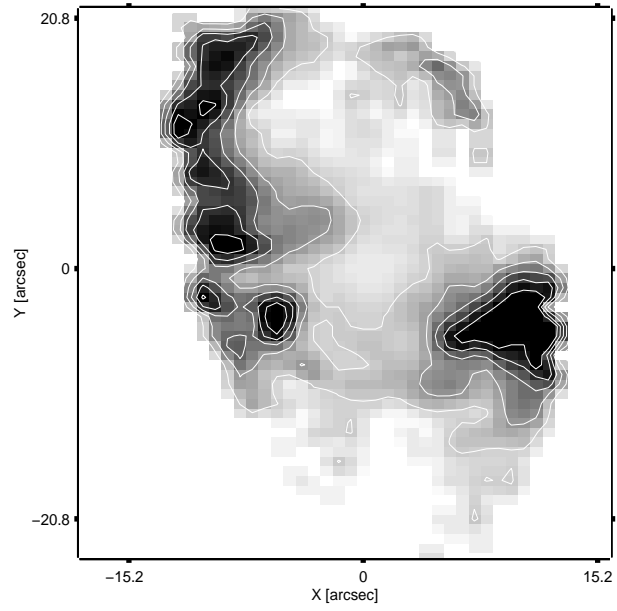,angle=0,height=3.2in,width=3.2in}
\caption{Intensity of the main gas component in (OIII [5007 \AA]) .}
\end{center}
\end{figure}

\begin{figure}
\begin{center}
\hspace{5cm}
\psfig{figure=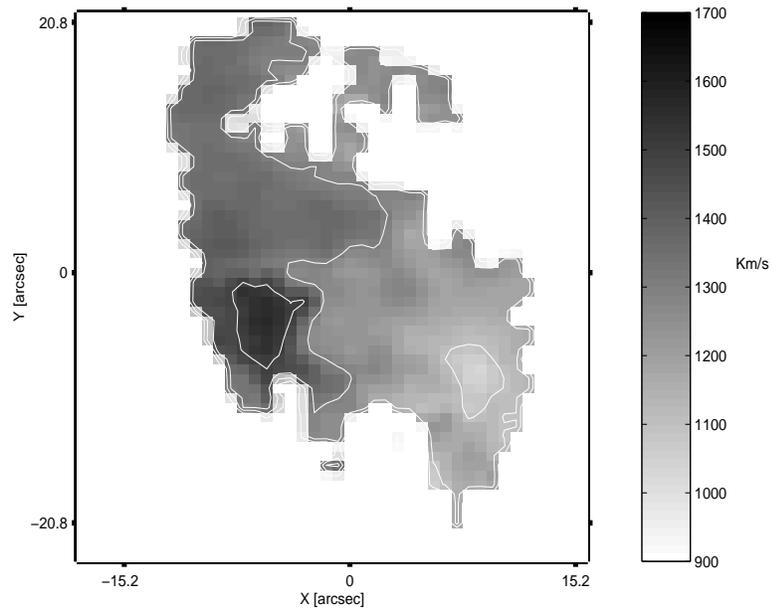,angle=0,height=3.2in,width=4in}
\caption{Velocity of the main gas component in (OIII [5007 \AA]).}
\end{center}
\end{figure}

\begin{figure}
\begin{center}
\hspace{-1cm}
\psfig{figure=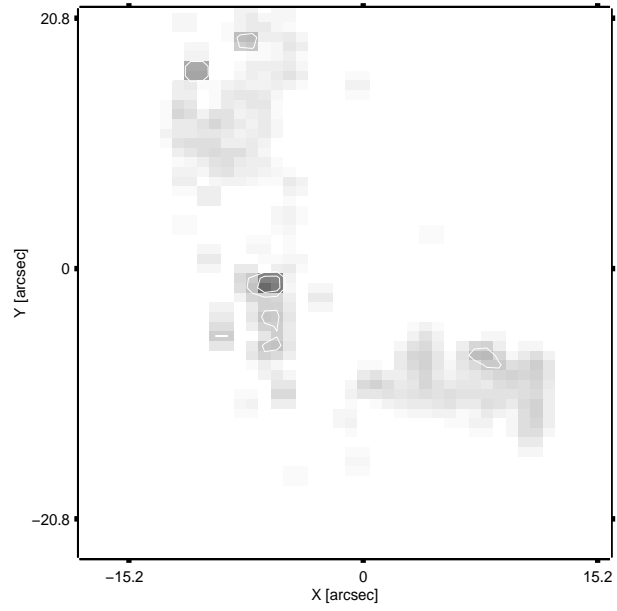,angle=0,height=3.2in,width=3.2in}
\caption{Intensity of the secondary gas component in (OIII [5007 \AA]).}
\end{center}
\end{figure}

\begin{figure}
\begin{center}
\hspace{5cm}
\psfig{figure=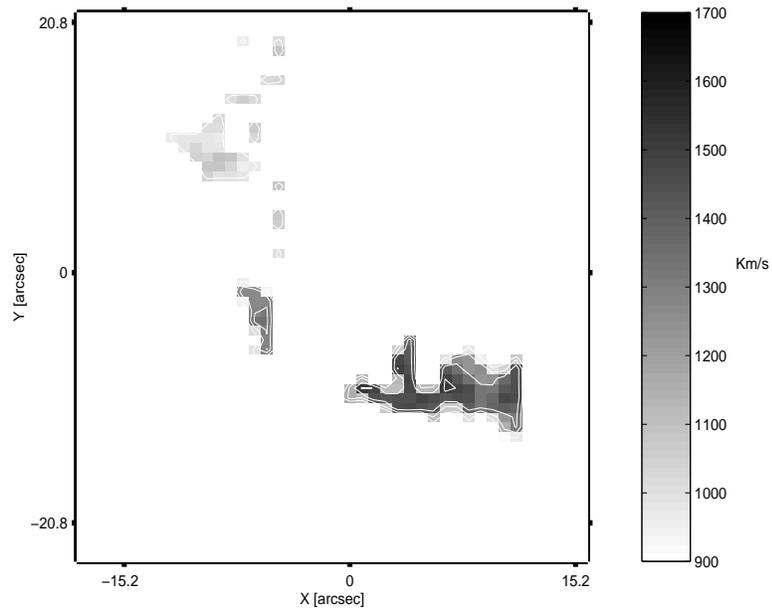,angle=0,height=3.2in,width=4in}
\caption{Velocity of the secondary gas component in (OIII [5007 \AA]).}
\end{center}
\end{figure}

\begin{figure}
\begin{center}
\hspace{5cm}
\psfig{figure=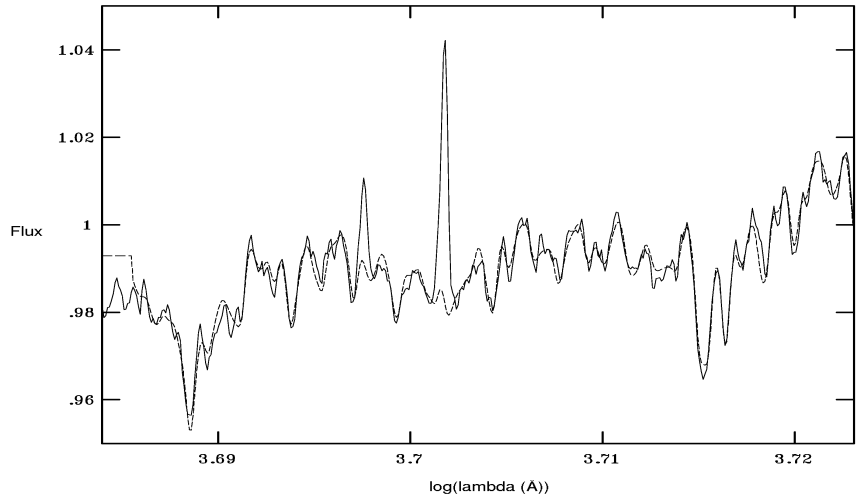,angle=0,height=2.6in,width=4.5in}
\caption{Synthethic stellar population models (dashed line) 
together with the central galaxy spectrum of NGC~7332 (solid line).}
\end{center}
\end{figure}

\begin{figure}
\begin{center}
\hspace{5cm}
\psfig{figure=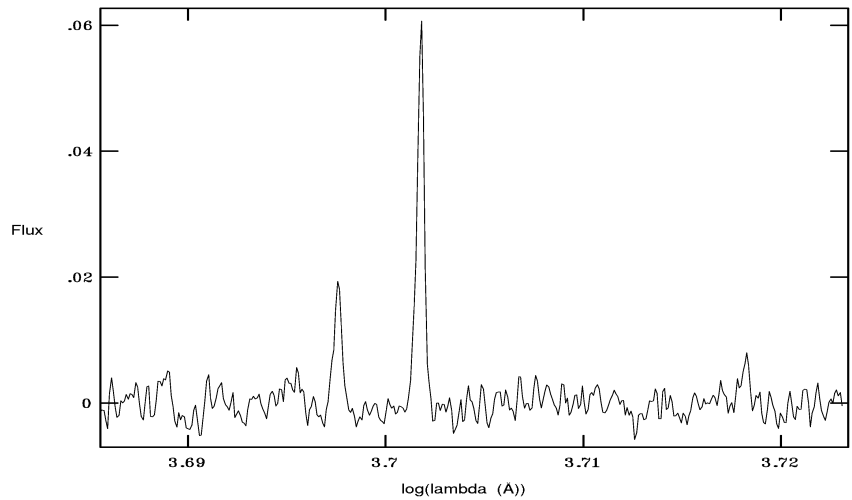,angle=0,height=2.6in,width=4.5in}
\caption{Residuals to the fit of Figure 8.}
\end{center}
\end{figure}

\end{document}